\documentclass[10pt]{iopart}
\usepackage{}
\usepackage{iopams} 
\usepackage{amssymb}
\usepackage{graphicx}
\usepackage{epstopdf}
\usepackage{bm}%
\usepackage{gensymb}
\usepackage{soul}
\usepackage{sidecap}
\usepackage[normalem]{ulem}
\usepackage[colorlinks,citecolor=blue,linkcolor=blue,anchorcolor=blue,filecolor=blue,urlcolor=blue]{hyperref}

\usepackage{epsfig}
\usepackage{graphicx}
\usepackage{amsfonts,amssymb}
\usepackage{color}
\usepackage {times}
\usepackage{epstopdf}

\newcommand{\bl}{\begin{aligned}}
\newcommand{\el}{\end{aligned}}

\newcommand{\be}{\begin{equation}}
\newcommand{\ee}{\end{equation}}   
\newcommand{\bea}{\begin{eqnarray}}
\newcommand{\eea}{\end{eqnarray}}
\newcommand{\ba}{\begin{array}}
\newcommand{\ea}{\end{array}}

\newcommand{\q}{{\bf q}}
\renewcommand{\k}{{\bf k}}
\newcommand{\Q}{{\bf Q}}

\graphicspath{{Figs/}}

\begin{document}

\title{Insight into interplay between bandstructure and Coulomb interaction via quasiparticle interference}

\author{Garima Goyal$^1$ and Dheeraj Kumar Singh$^1$}

\address{ Department of Physics and Material Science, Thapar Institute of Engineering and Technology, Patiala-147004, Punjab, India}     
\date{\today}

\ead{dheeraj.kumar@thapar.edu}
\begin{abstract}
Quasiparticle interference has been used frequently for the purpose of unraveling the electronic states in the vicinity of the Fermi level as well as the nature of superconducting gap in the unconventional superconductors. Using the metallic spin-density wave state of iron pnictides as an example, we demonstrate that the quasiparticle interference can also be used as a probe to provide crucial insight into the interplay of the electronic bandstructure and correlation effects in addition to bringing forth the essential features of electronic states in the vicinity of the Fermi level. Our study reveals that the features of quasiparticle interference pattern can help us narrowing down the interaction parameter window and choose a more realistic tight-binding model.

\end{abstract}
\pacs{}
\maketitle
\ioptwocol
\newpage
\section{Introduction}
 The scanning-tunneling microscopy (STM), which can be used to map out the quasiparticle interference (QPI) patterns generated by the impurity atoms, is one of the most powerful experimental techniques used for unraveling of the nature of electronic states in the vicinity of the Fermi surface~\cite{hoffman, petersen, fujita, mcelroy, capriotti, kriesel}. The information about the electronic state, thus obtained, can provide us with important insight into the origin of different phases exhibited by a material system as a function of temperature and carrier doping as well as their unique behavior~\cite{christianson, boker, lee, chen1, andersen, paul, wang, hoffman1}.

The impurity atom, along the surface scanned by the STM, scatters off the quasiparticles elastically from an electronic state with momentum ${\k_1}$ to another state with momentum ${\k_2}$, with both states lying along the constant-energy contours (CEC). Consequently, there is a spatial modulation in the local density of states (LDOS) referred to as QPI. The modulation wavevector ${\bf q} = {\bf k}_2 - {\bf k}_1$~\cite{crommie,hoffman} has often been used to decode the nature of the Fermi surfaces. It may be noted that the QPI patterns have also been used to unravel the nature of superconducting order parameter~\cite{sykora, ding, hirschfeld}. For instance, QPI patterns generated by the magnetic and non-magnetic impurity have been used to distinguish between the sign-changing and sign-preserving superconducting order parameters~\cite{barnea, maltseva, hanaguri, akbari, dheeraj, dheeraj1,rashid, chi, chi1}.

STM has been used extensively in studying the QPI patterns in the iron-based unconventional superconductors in a variety of phases including the spin-density wave (SDW) state~\cite{knolle}, sign-changing $s$-wave superconducting state~\cite{akbari, zhang1, hanaguri}, orthorhombic nematic phase~\cite{chuang, rosenthal}, etc. The QPI patterns consist of highly-anisotropic nanostructures across different phases while the length scale of the modulation was suggested to be material dependent~\cite{allan, knolle, barnea, maltseva, sykora, ding}. Moderate anisotropic LDOS modulation is not surprising for the SDW state with ordering wavevector ($\pi, 0$), as the four-fold rotation symmetry is broken~\cite{gastiasro, dheeraj2}. However, the existence of such structures even in the high-temperature tetragonal phase or in the superconducting state has been puzzling~\cite{walker}. Theoretical studies indicate that such anisotropic features in the QPI patterns can be explained provided that non-vanishing orbital splitting between the $d_{xz}$ and $d_{yz}$ orbitals is assumed to exist though the origin of the same remains unclear~\cite{inoue, eremin, plonka, daghofer}.
Majority of the theoretical works investigating QPI patterns in the iron-based superconductors (IBS) considered only the intraorbital scattering, i.e., preserved orbital state upon scattering. Although, these works did not reflect on how a specific orbital-distribution along the Fermi pockets will affect the patterns.

IBS are considered to be moderately correlated electron systems~\cite{nakamura, miyake, anisimov, kurmaev} as various estimates put the largest interaction, i. e., intraorbital Coulomb interaction $U$ to be near $\sim $1eV while the bandwidth suggested by the bandstructure calculations lies near $\sim$ 4eV~\cite{skornyakov, yang, vildosola, anisimov1, anisimov2, shorikov, kroll}. Second, all the five 3$d$ orbitals of iron atom contribute at the Fermi level though the dominant contributions come only from $d_{xz}$, $d_{yz}$, and $d_{xy}$ orbitals~\cite{graser, kuroki, eschrig}. Interestingly, the latter orbitals continue to contribute at the Fermi level even in the SDW state of iron pnictides, despite that it is accompanied by a magnetic order with broken four-fold rotation symmetry. This follows from an earlier prediction which points out that the SDW state should be metallic, a simple consequence of symmetry preserved existence of Dirac cone~\cite{ran, yu}. Thus, several bands cross the Fermi level giving rise to Fermi pockets dominated by the $d_{xz}$, $d_{yz}$, and $d_{xy}$~\cite{dheeraj3, yi1}. The evidence of these pockets can directly be obtained via angle-resolved photoemission spectroscopy (ARPES)~\cite{yin1, yi, richard, liu2} or indirectly through the STM measurements~\cite{yin2}.

{\begin{table*}[hbtp]
\scriptsize
	\resizebox{\textwidth}{!}{%
		\begin{tabular}{|p{0.14\linewidth}|p{0.18\linewidth}|p{0.18\linewidth}|p{0.18\linewidth}|}
			\hline
	        {\bf Feature}  &  {\bf Graser et al.}  &  {\bf Ikeda et al.}  &  {\bf Kuroki et al.}  \tabularnewline
			\hline
			{\bf Ab-initio method}  &  Structural relaxation, optimized crystallographic parameters &  Experimental crystallographic parameters  &  Experimental crystallographic parameters  \tabularnewline
			\hline

			\hline
			{\bf Unit cell}  &  Fe-Fe plane in which a single Fe forms a unit cell  &  FeAs plane with two Fe atoms per unit cell  &  FeAs plane with two Fe atoms per unit cell  \tabularnewline
			\hline
            {\bf Electron pockets at $\bm{(\pi, 0)}$}  &   Moderate ellipticity  &  Moderate ellipticity  &   Nearly circular  \tabularnewline
			\hline
			{\bf Hole pocket at $\bm{(\pi, \pi)}$}  &  No pocket  &  Pocket of largely $d_{xy}$ character  &   Pocket of largely $d_{xy}$ character \tabularnewline
			\hline
			{\bf Hole pockets at $\bm{(0, 0)}$}  &  Squircle shaped & Squircle shaped  &   Circle shaped  \tabularnewline
			\hline
			{\bf Nesting between electron and hole pockets}  &   Weaker as overlapping is not good, weaker tendency towards ($\pi, 0$) magnetic order  &   Relatively better because of larger overlap, better tendency towards ($\pi, 0$) magnetic order &  Relatively better because of larger overlap, better tendency towards ($\pi, 0$) magnetic order  \tabularnewline
			\hline
		\end{tabular}}
        \caption{Key differences between five-orbital tight-binding  models due to Graser \textit{et al.}, Ikeda \textit{et al.}, and Kuroki \textit{et al.}} \label{t1}
\end{table*}
There are several questions of strong interest with regard to the interplay of bandstructure and correlation effects. This is because not only the shape and size of Fermi pockets may depend on this interplay but also the orbital-weight redistribution along the reconstructed Fermi surface. In particular, how much is the shape or size of the individual Fermi pockets sensitive to a small change in the largest interaction parameter $U$? Is the change with $U$ also dependent on a particular tight-binding model? These questions are of significant importance especially in the four-fold symmetry broken SDW state. This is largely due to the fact that an anisotropic shape and orbital distribution along the Fermi surfaces will give rise to anisotropic scattering of the quasiparticle leading to anisotropy in various electronic properties. Therefore, a better understanding of the nature of QPI may enable us to restrict the interaction parameter window as well as to choose a more realistic tight-binding model.

In order to address the issues mentioned above, in this paper, we examine three frequently used five-orbital tight-binding models, namely due to Graser \textit{et al.}~\cite{graser}, Ikeda \textit{et al.}~\cite{ikeda}, and Kuroki \textit{et al.}~\cite{kuroki}. All of them are obtained by fit to the ab-initio band structures of LaFeAsO. The key differences in these models are listed in the Table 1.
We demonstrate how the QPI patterns can prove to be a useful tool to get important insight into the interplay of details of electronic states including the orbital distribution in the vicinity of Fermi level and electronic correlations. More specificially, an aspect unexplored earlier, we find that the orbital-weight distribution along the Fermi pockets is very sensitive to change in $U$. The degree of sensitivity is the largest in the model due to Graser \textit{et al.} and the least in the model due to Ikeda \textit{et al.}. An immediate consequence is that the QPI patterns show a comparatively quicker changes in the model due to Graser \textit{et al.} as $U$ changes. The LDOS modulation of $\sim 8a$, with $a$ as lattice constant, in the model due to Ikeda \textit{et al.} agrees qualitatively with the experiments, and it is robust against change in $U$. We find that none of the other remaining models show a qualitative agreement with the experiments, besides being non robust with change in $U$. Therefore, the current work highlights the limitation of the intensively used model of iron pnictides in explaining the existence of nearly one-dimensional modulation in LDOS as observed in STM measurements. In all the three models, the existence of a large hole pocket around $\Gamma$ prove to be detrimental to the nearly one-dimensional LDOS modulation, which can potentially arise from the scattering between the Dirac pockets.

This manuscript is organized into several sections. Section II describes the tight-binding Hamiltonian with on-site Coulomb interaction for multiorbital models and the self-consistent mean-field methodology used to obtain the decoupled Hamiltonian in the SDW state. It then details the treatment of single-particle impurity-induced scattering within the framework of $T$-matrix approximation. Section III presents the results of QPIs obtained for different models for various parameter sets. Finally, Section IV concludes the manuscript with discussions.

\section{Model and method}

\subsection{Multi-orbital models of iron pnictides}
In order to study the QPI patterns in the SDW state, we employ the following multi-orbital tight-binding part of the Hamiltonian
 \begin{equation}
 \mathcal{H}_k  =  \sum_{\bf{k}} \sum_{\alpha ,\beta} \sum_{\sigma} \varepsilon_{\bf{k}}^{\alpha \beta} \textit{d}_{\bf{k} \alpha \sigma}^{\dagger}  \textit{d}_{\bf{k} \beta \sigma},
 \end{equation}
where $\textit{d}_{\bf{k}\alpha\sigma}^{\dagger} (\textit{d}_{\bf{k}\alpha\sigma})$
is the creation (destruction) operator for an electron with spin $\sigma$ and momentum ${\bf k}$ in the orbital $\alpha$, where $\alpha/\beta$ denotes the five 3$d$ orbitals of the iron atom. The matrix elements $\varepsilon_{\bf{k}}^{\alpha \beta}$ contains information about both the momentum dependent intra and interorbital hopping energies as well as on-site energies, which are taken from the Refs.~\cite{graser, ikeda, kuroki} depending on the model being used.

The interaction Hamiltonian includes the standard on-site Coulomb repulsion terms given by
 \begin{eqnarray}
 \mathcal{H}_{int}  =  \textit{U} \sum_{{\bf i}  \alpha} \textit{n}_{\bf{i} \alpha \uparrow} \textit{n}_{\bf{i}  \alpha \downarrow}  +  (\textit{U}^{\prime}  -  \frac{\textit{J}}{2}) \sum_{\bf{i}, \alpha < \beta} \textit{n}_{\bf{i} \alpha} \textit{n}_{\bf{i} \beta}  - \nonumber \\
 2 \textit{J} \sum_{\bf{i}, \alpha < \beta} \textit{\bf S}_{i \alpha} \cdot \textit{\bf S}_{i\beta} +  \textit{J} \sum_{\bf{i}  ,  \alpha < \beta  ,  \sigma} \textit{d}^{\dagger}_{{\bf i}\alpha \sigma} \textit{d}^{\dagger}_{{\bf i}\alpha \bar{\sigma}}\textit{d}_{{\bf i}\beta \bar{\sigma}} \textit{d}_{{\bf i}\beta \sigma} 
 \end{eqnarray}
The first and second terms describe intra- and interorbital Coulomb interaction respectively, where ${n}_{{\bf i}\alpha \sigma} = d^{\dagger}_{{\bf i} \alpha \sigma} d_{{\bf i} \alpha \sigma}$ is the number operator for the electrons of spin $\sigma$ at site {\bf i} in an orbital $\alpha$.  The third and fourth terms represent Hund’s coupling and pair hopping, respectively, where $\bar{\sigma}$ denotes the spin anti-parallel to $\sigma$. The relation $\textit{U}   =  \textit{U}^{\prime}  + 2 \textit{J}$ is ensured to maintain the rotational symmetry. Throughout, we set $J = 0.25U$ as indicated by various estimates~\cite{yang, aichhorn, miyake} whereas all the energy parameters are in units eV unless stated otherwise.
 
\subsection{Mean-field Treatment of Hubbard term}
The interaction part of the Hamiltonian consists of terms quartic in the electron creation and annihilation operators. Using meanfield decoupling, the quartic terms can be decoupled into terms bilinear in the electron-field operators for the SDW state with ordering wave vector $\Q = (\pi,0)$. For simplicity, we assume that the spins are pointing along the $z$ direction without any loss of generality, therefore, the resulting Hamiltonian is
\begin{eqnarray}    \label{eqn:6}
\mathcal{H}_{mf}  &=&  \sum_{\bf{k} \sigma} \Phi^{\dagger}_{{\bf k} \sigma}
\left(\begin{array}{cc}\hat{\varepsilon}_{\bf k}+ \hat{N} & \hat{\Delta} \\ \hat{\Delta} &  \hat{\varepsilon}_{\bf k+Q} + \hat{N} \end{array}\right)
\Phi_{{\k} \sigma} \nonumber\\
&=& \sum_{\bf{k} \sigma} \Phi^{\dagger}_{\bf{k} \sigma}
\hat{H}_{SDW}(\k)
\Phi_{{\k} \sigma},
\end{eqnarray} 
\noindent where $\Phi^{\dagger}_{\bf{k} \sigma}$  =  $(\textit{d}^{\dagger}_{{\k} 1 \sigma} \cdots \textit{d}^{\dagger}_{{\bf k} 5 \sigma}, \textit{d}^{\dagger}_{(\bf{k + Q}) 1 \sigma} \cdots
\textit{d}^{\dagger}_{{(\k+\Q)} 5 \sigma})$ and $\hat{H}_{SDW}$ is the Hamiltonian matrix in the SDW state. Each matrix element in the above Hamiltonian is itself a 5 $\times$ 5 matrix. The indices 1, 2, 3 etc. correspond to the 3$d$ orbitals of iron atom, i.e., $d_{xz}$, $d_{yz}$, $d_{xy}$, $d_{x^2 - y^2}$, and $d_{3z^2 - r^2}$. The exchange field operators $\hat{\Delta}$ and $\hat{\textit{N}}$ can be written in terms of on-site interaction parameters, orbital magnetization, and orbital-resolved charge densities as
\begin{eqnarray}
  2  \Delta_{\alpha \alpha}  &=&  \textit{U} \textit{m}_{\alpha \alpha}  +  \textit{J} \sum_{\alpha \neq \beta} \textit{m}_{\beta \beta}   \label{eqn:7}  \nonumber  \\
  2 \textit{N}_{\alpha \alpha} &=& \textit{U} n_{\alpha \alpha} + (\textit{2U - 5J}) \sum_{\alpha \ne \beta} n_{\beta \beta} \nonumber \\
  &=&  \textit{(5J - U)} \textit{n}_{\alpha \alpha},
\end{eqnarray}
where we have used  $\sum_{\alpha} {n}_{\alpha \alpha} = c$ with total band filling being $c = n_{tot}  = 6$ that is fixed throughout.  The order parameters $m_{\alpha \alpha /\beta \beta}$ and $n_{\alpha \alpha /\beta \beta}$ given by
\begin{eqnarray}
 n_{\alpha \alpha} &=& \sum_{\k \sigma} \langle d^{\dagger}_{\k \alpha \sigma} d^{}_{\k \alpha \sigma}\rangle \nonumber\\
m_{\alpha \alpha} &=& \sum_{\k \sigma} \langle d^{\dagger}_{\k+\Q \alpha \sigma} d^{}_{\k \alpha \sigma}\rangle
\end{eqnarray}
are calculated self-consistently using eigenvalues and eigenvectors of 10 $\times$ 10 matrix Hamiltonian $\hat{H}_{SDW}$ for the SDW state. The Hamiltonian is diagonalized via standard subroutine for a Hermitian matrix available via LAPACK. The self-consistent procedure begins with some initial value of orbital-resolved magnetization. Thereafter, the eigenvalues and eigenvectors of $\hat{H}_{SDW}$ are used in Eq. 5 to calculate the new orbital-resolved magnetization and charge density, which are, then, used in the next cycle of self-consistency scheme. The producedure is repeated until the convergence in the magnetization for each of the orbitals is reached at the $ni^{th}$ iternation,  i. e., $\delta m =  m_{ni} -  m_{ni-1} < \epsilon$. In the current work, we have used $\epsilon \sim 10^{-6}$.  \\

\subsection{T-matrix approximation}
QPI is studied by calculating the modulation in Green's function due to impurity-induced scattering. For this, we adopt Green's function formalism to obtain the modulation in the electronic DOS. We consider a single non-magnetic  impurity with $\delta$-potential such that the impurity matrix is momentum independent, $V_{1}({\k_{\bf1}, \k_{\bf 2}}) = V$. For simplicity, we focus on those type of impurities, which preserve the orbital states of the scattered quasiparticle. The STM tip used currently are not spin or orbital sensitive, and therefore they cannot provide information about the spin or orbital state of the scattered quasiparticle.}

The bare Green's function in the $(\pi,0)$-type SDW state is given by
\begin{equation}
     \hat{G}^{0} ({\bf k}, \omega) = ((\omega + i \eta) \hat{I} - {\hat{H}}_{SDW}(\k))^{-1} \nonumber
\end{equation}
where $\hat{I}$ is a $10 \times 10$ identity matrix and $\eta$ is infinitesimally small positive number ($\sim$ 0.005) used as a broadening parameter.

The change induced in the Green's function due to scattering by an impurity is calculated using $T-$matrix approximation~\cite{zhang1, kriesel1, choubey}, i.e.,
\begin{equation}
    \centering
      \delta \hat{G} ({\bf k, k+q}, \omega) = \hat{G}^{0} ({\bf k}, \omega) \hat{T} (\omega) \hat{G}^{0} ({\bf k + q}, \omega)
\end{equation}
where $\hat{T}$ matrix is defined as
\begin{equation}
    \centering
       \hat{T} (\omega) = (\hat{{I}} - \hat{\mathcal{G}}^{0} (\omega))^{-1} \hat{V}_{imp}.
\end{equation}
The element of matrix $\hat{\mathcal{G}}^{0} (\omega)$ is obtained via summing over all the momenta for each of the elements of the $\hat{G}^{0} ({\bf k},\omega)$ matrix, \textit{i.e.},
\begin{equation}
    \centering
       \hat{\mathcal{G}}^{0} (\omega) = \frac{1}{N} \sum_{\k} \hat{G}^{0} ({\bf k}, \omega).
\end{equation}
$\hat{V}_{imp}$ is given by
\begin{equation}
    \centering
       \hat{V}_{imp} = {\hat{V}}_{0}
    \left(\begin{array}{cc} \hat{I} & \hat{I} \\ \hat{I} & \hat{I} \end{array}\right)
\nonumber,
\end{equation} 
which denotes the $10 \times 10$ scattering matrix due to an impurity atom, where $\hat{I}$ is the $5 \times 5$ identity matrix. Diagonal matrices describes normal intraorbital scattering whereas the off-diagonal matrices corresponds to the intraorbital Umklapp scattering. The matrix inverse in Eq. 7 is obtained by using the standard subroutine for the inversion of general complex matrix available via LAPACK.

Next, the modulation $\delta \rho ({\bf q}, \omega)$ induced in DOS due to the scattering by the impurity atom can be obtained in terms of the modulation in the Green's function given by
\begin{eqnarray}
    \centering
       \delta \rho ({\bf q}, \omega) &=& -\frac{i}{2 \pi} \sum_{\bf k} g({\k}, {\q}, \omega),
\end{eqnarray}
where
\begin{equation}
    g({\k}, {\q}, \omega) = Tr \delta \hat{G}({\k}, {\k+\q}, \omega) - Tr \delta \hat{G}({\k+\q}, {\k}, \omega) \nonumber.
\end{equation}
The real space QPI pattern can be obtained by
\begin{equation}
    \delta \rho ({\bf r}, \omega) = \frac{1}{N} \sum_{\bf k} \delta \rho ({\bf q}, \omega) e^{i {\bf q} \cdot {\bf r}}.
\end{equation}

\section{Results}
\subsection{Five-orbital model of Ikeda \textit{et al.}}

The five-orbital model due to Ikeda \textit{et al.}~\cite{ikeda} is one of the widely used tight-binding models of iron pnictides that includes all the five $3d$ orbitals of iron. It is based on the \textit{ab-initio} method and fluctuation-exchange approximation (FLEX) has been used within this model to investigate the superconducting instability. Proposed originally in the folded-Brillouin zone, the axes are rotated by $\pi/4$ with respect to the lattice formed by iron atoms to accommodate two atoms per unit cell. In the following, we focus on the unfolded Brillouin zone containing only a single iron atom per unit cell.

\begin{figure}[hb]
    \centering
    \includegraphics[scale = 1.0, width = 8cm]{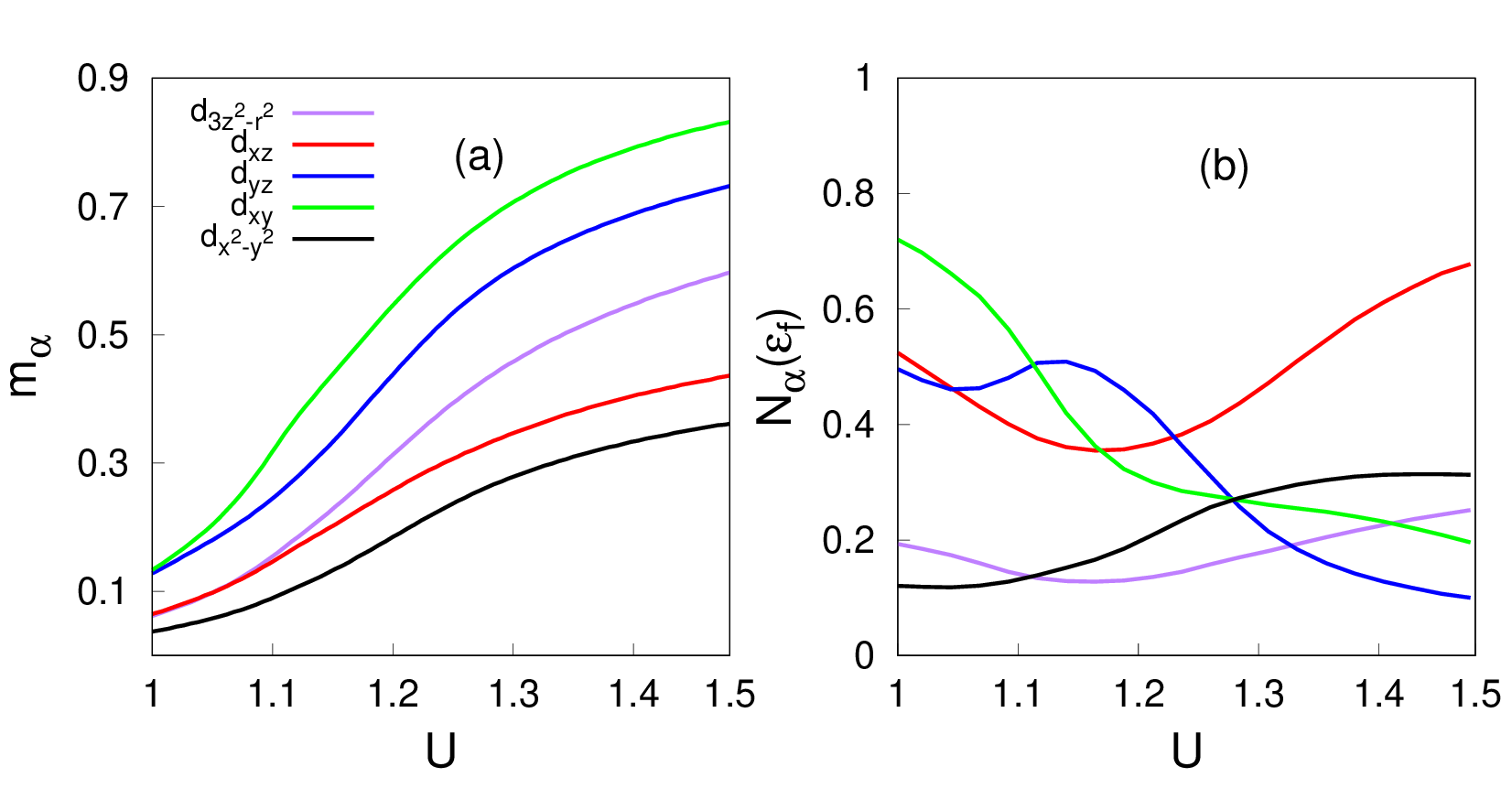}
    \caption{(a) Orbital-resolved magnetization and (b) DOS at the Fermi level as a function of $U$ in the SDW state of the five-orbital model of Ikeda \textit{et al.}}
    \label{1}
\end{figure}

Fig~\ref{1}(a) shows self-consistently obtained orbital magnetization parameter as a function of $U$. The magnetization of each orbital increases monotonically with $U$, however, a slightly faster rise around $U \sim 1.2$ may be noticed. In particular, orbital magnetizations $m_{xy}$ and $m_{yz}$ approach $\sim$ 1 with further increase in $U$ indicating that the orbital-charge density in these two orbitals may be drawing close to the half filling. Beyond  $U \sim 1.2$, the magnetization of individual orbitals changes slowly and total magnetization is significantly larger in comparison to 1, which does not correspond to the experimentally observed magnetic moments even if the quantum corrections to the sublattice magnetization are incorporated ~\cite{cruz, si, klauss, wu}.

Figure \ref{1}(b) shows the orbital-resolved DOS at the Fermi surface as a function of $U$. For smaller $U$, the $d_{xy}$ orbital is predominant, with $d_{xz}$ and $d_{yz}$ orbitals being the other major contributors. $d_{xy}$-orbital weight continuously declines as $U$ increases. The $d_{xz}$-orbital weight also shows a dip, while the weight corresponding to $d_{yz}$ orbital peaks around $U \sim 1.2$, afterwards, they increase and decrease in a monotonic way, respectively, as $U$ continues to increase. Suppression of orbital weight of $d_{yz}$ and $d_{xy}$ orbitals for higher $U$ is indicative of orbital-selective insulating tendency as also reflected in magnetization of these orbitals approaching unity noted earlier. For smaller $U$, $d_{3z^2-r^2}$- and $d_{x^2-y^2}$-orbital weights are small. This suggests that the redistribution of orbital weights is highly sensitive to the choice of $U$, which can be very helpful in understanding the QPI patterns.

\begin{figure} [hb]
    \centering
    \includegraphics[scale = 1.0, width = 8cm]{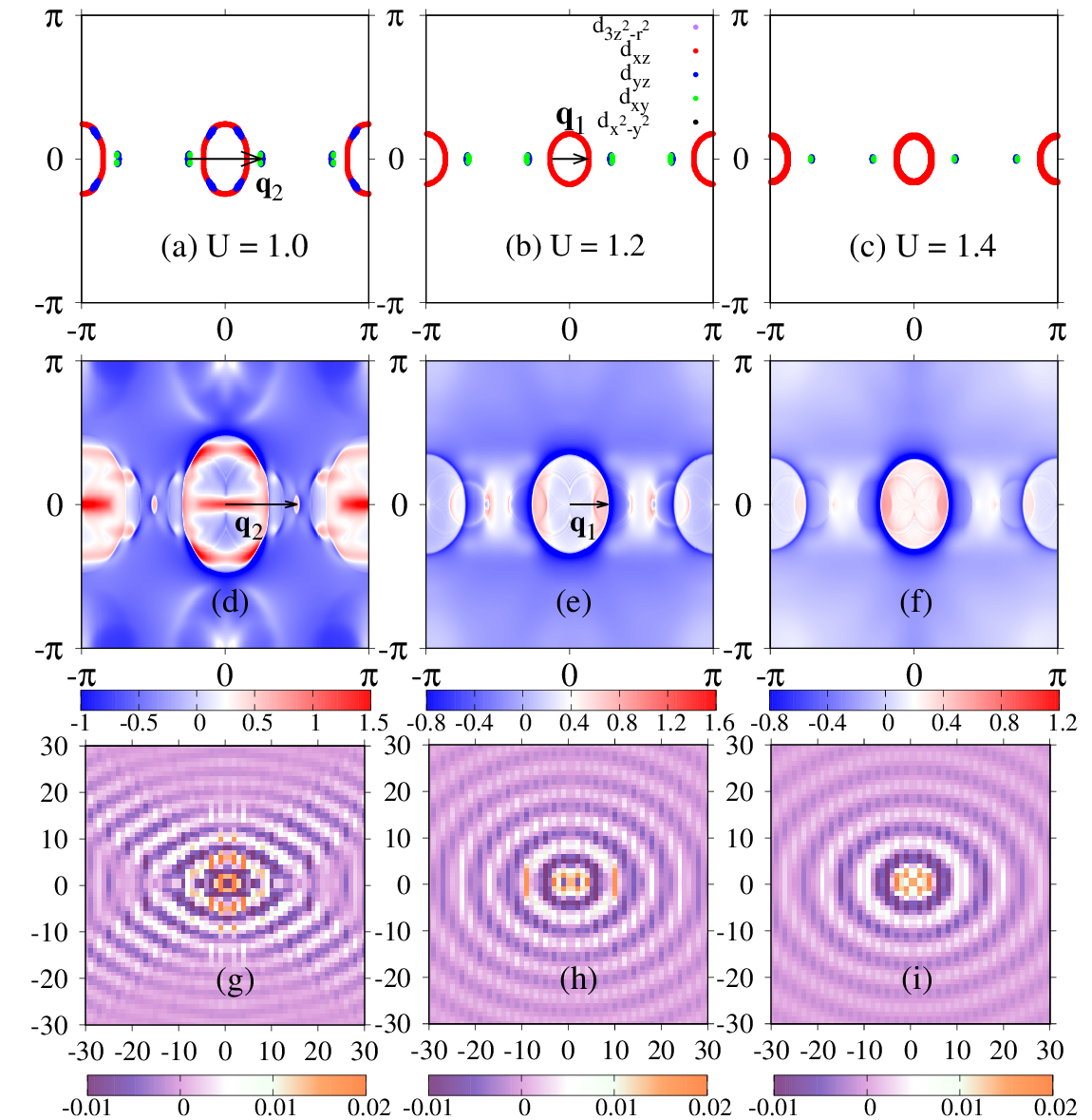}
    \caption{The three horizontal panels, starting from the top to bottom display CECs with dominant-orbital contributions, associated $q$-space, and $r$-space QPI patterns, respectively, in the SDW state of the five-orbital model of Ikeda \textit{et al.} for $\omega = -20$ meV whereas $U$ varies. }
    \label{2}
\end{figure}
Figure \ref{2}(a-c) show the constant-energy contours (CEC) with orbital-weight distributions all along for $\omega = -20$ meV for different values of $U$. The Fermi pockets show robustness in their shape, structure, and orbital-weight distribution against any change in $U$ for the range $1.0 \lesssim U \lesssim 1.4$, beyond which, the total magnetization is too large to be compared to experimental values. For all the three cases, there exists a large and nearly elliptical hole pocket at $\Gamma$ with a dominant $d_{xz}$ character, accompanied by a pair of tiny hole pockets of mixed $d_{yz}$ and $d_{xy}$ characters located around ${\k} \sim (\pm 0.25 \pi,0)$. Here, it may be noted that even a difference of 1\% in the orbital contributions may indicate the dominance of particular orbital according to the coloring scheme of the CEC adopted in the current work. This is especially true near transition points, where the dominance of orbital changes. As $U$ increases, the size of these pockets decreases but only slightly. Since only intraorbital scattering is allowed, the orbital distribution of CECs permit mainly two types of predominant scattering, one is the intrapocket scattering vector ${\q}_1$ connecting different regions of the hole pocket around $\Gamma$, which is dominated by the $d_{xz}$ orbital. The second scattering vector is ${\q}_2$, which connects two tiny CECs dominated by $d_{yz}$ and $d_{xy}$. Intrapocket scattering vector ${\q}_1 \sim (0.25 \pi, 0)$ generates an elliptical pattern in momentum space (Fig.~\ref{2}(d-f)) so that the real-space modulation vector $R \sim$ 8$a$ along $x$ direction (Fig.~\ref{2}(g-i)). ${\q}_2$ has the potential to produce strong quasiparticle peaks at $q_x \sim \pm 0.5 \pi$, resulting into a nearly one-dimensional modulation along $x$ direction. However, for $U = 1.0$, the central elliptical pocket has some flavor of $d_{yz}$ along the diagonal direction indicating a larger spectral weight, which leads to a strong modulation in the diagonal direction. On the other hand, along the $x$ direction, an overlap between the patterns generated by ${\bf q}_1$ and ${\bf q}_2$ appears to destroy the expected one-dimensional modulation in real space (Fig.~\ref{2}(g-i)). However, as $U$ increases, the $d_{xy}$-orbital weight decreases, the tiny pockets become smaller and smaller so the pattern generated by ${\bf q}_2$ becomes weaker and the final pattern appears dominated by the scattering vector ${\bf q}_1$.

The interpocket scattering between one of the tiny CECs and the elliptical hole pocket around $\Gamma$ is not able to generate a noticeable pattern for the reason that they are dominated by different orbitals. However, if interorbital scattering is allowed, this can potentially yield a nearly one dimensional modulation.

\subsection{Five orbital model of Graser \textit{et al.}}

The five-orbital model due to Graser \textit{et al.}~\cite{graser} is another frequently used tight-binding model for the iron pnictides. This model was constructed by fitting the hopping parameters with the bandstructure obtained via first-principles for LaOFeAs and it uses an unrotated orbital basis of the Fe-Fe lattice plane, with the crystallographic unit cell consisting of a single Fe atom.

\begin{figure} [b]
    \centering
    \includegraphics[scale = 1.0, width = 8cm]{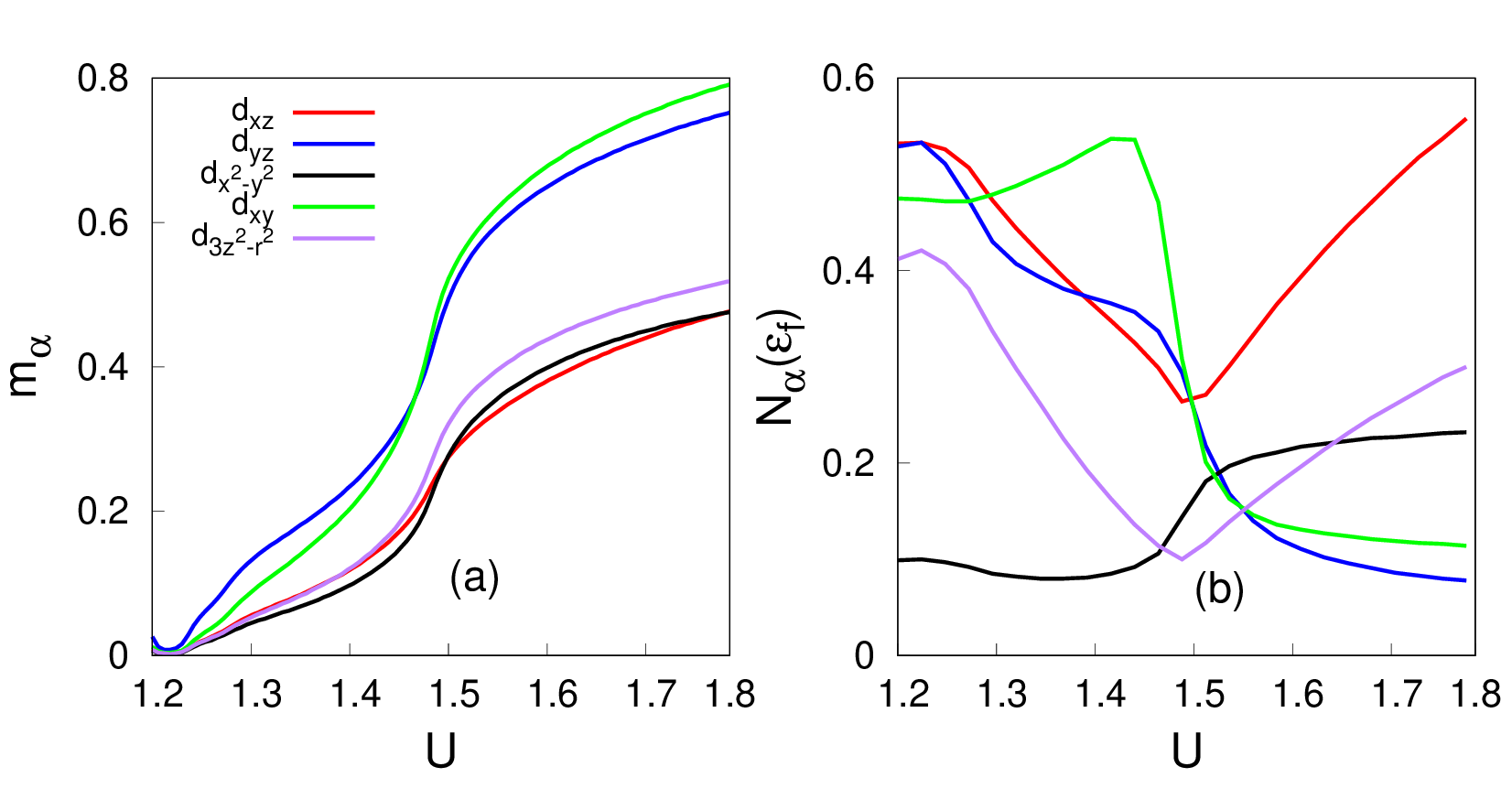}
    \caption{(a) Orbital magnetization and (b) DOS at the Fermi level as a function of $U$ in the $(\pi,0)$ SDW state within the five-orbital model of Graser \textit{et al.}}
    \label{3}
\end{figure}

Figure~\ref{3}(a) shows the orbital-resolved magnetization and DOS at the Fermi level as a function of $U$. $U_c$ needed for the onset of SDW state is slightly larger than that in the model due to Ikeda \textit{et al}. As expected, the magnetization for each of the orbitals increases monotonically with $U$ once a critical $U_c \sim 1.2$ corresponding to the onset of SDW state is crossed. There is another sharp jump in the magnetic moments of all the orbitals near $U \sim 1.5$, driven by the tendency towards orbital-selective Mott transition and Hund's coupling ~\cite{yu, medici, ishida}. Although, beyond $U \sim 1.5$, the total magnetization becomes significantly larger than 1. Therefore, we consider $1.3 - 1.7$, a more appropriate range for $U$, to study the QPI patterns. In this weak-to-intermediate coupling regime, the Fermi pockets in the metallic SDW state may be very sensitive to any change in $U$, which can be captured through QPI patterns.

Figure \ref{3}(b) presents the orbital-resolved DOS at the Fermi surface. For smaller values of $U$, the DOS is nearly same for the  dominant $d_{xz}$ and $d_{yz}$ orbitals. Even though $U$ is increased until $\sim 1.5$eV, the DOS continues to be nearly same for both the orbitals, while they decrease continuously in a monotonic way. Beyond this point, the DOS of $d_{xz}$ orbital starts to rise beyond $U \sim 1.5$, whereas $d_{yz}$ continues to decrease, indicating an increased tendency towards orbital ordering. Moreover, the orbital-charge density for the $d_{yz}$ orbital appears to move towards half filling, $n_{yz} = 1$ prompting an orbital-selective insulating behavior noted earlier. On the other hand, the DOS for $d_{xy}$ orbital rises and dominates in the range 1.3 - 1.5, afterward it also shows a sharp decline. This fact together with continuous increase in magnetic moment indicates that $n_{xy}$ also approaches 1. Beyond $U \sim 1.5$, it is the DOS of $d_{xz}$ which dominates. The non-dominant orbitals such as $d_{3z^2 - r^2}$ orbital follows the same pattern as $d_{xz}$ while $d_{x^2 - y^2}$ orbital has the least contribution.

\begin{figure}[tb]
    \centering
    \includegraphics[scale = 1.0, width = 8cm]{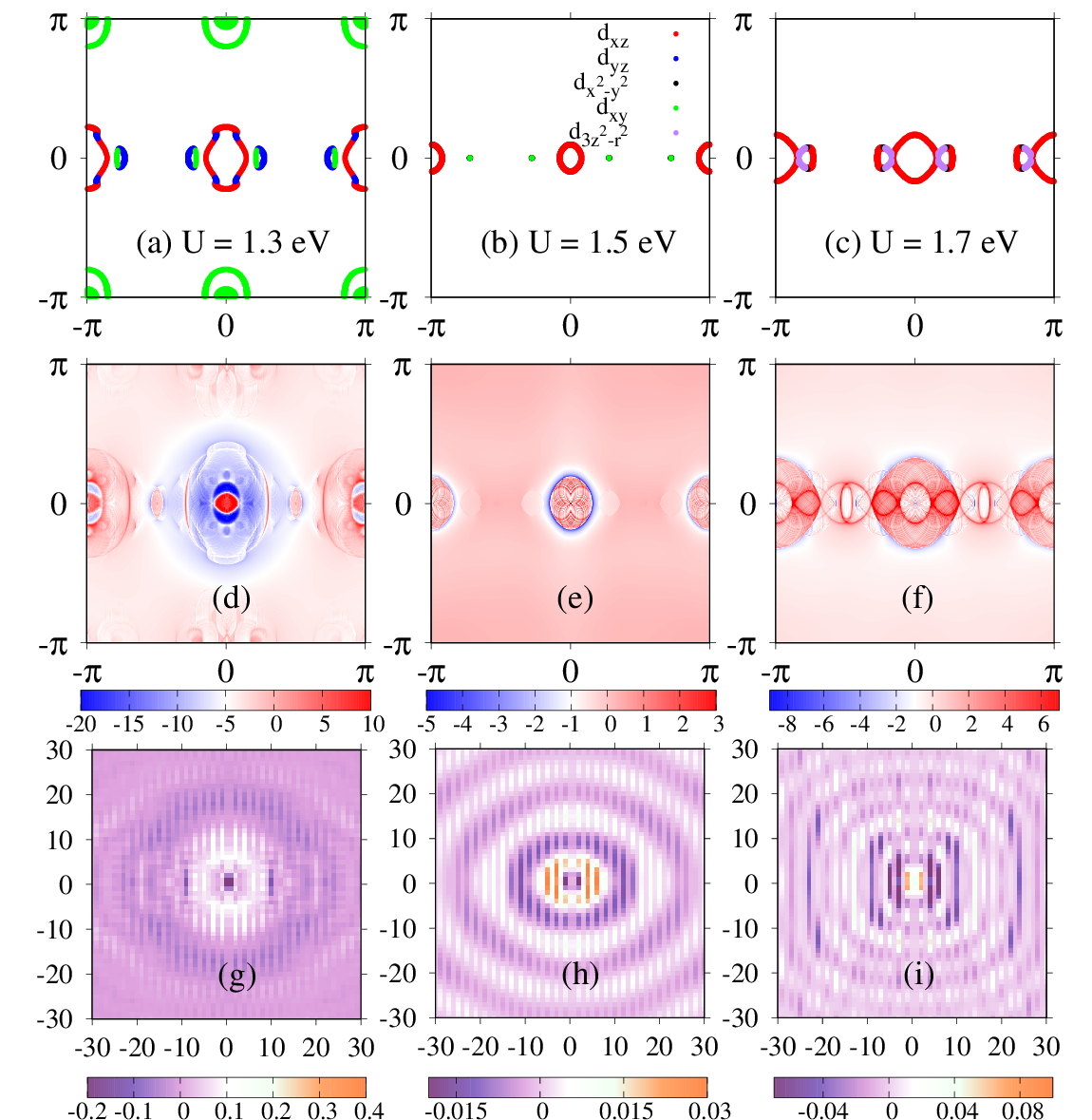}
    \caption{ Each column contains CECs (a-c), $q-$ space (d-f) and $r-$ space (g-i) QPI maps for $U = 1.3$, $1.5$, and $U = 1.7$ eV, respectively at $\omega$ = -20 meV in the SDW state within the five-orbital model of Graser \textit{et al.}}
    \label{4}
\end{figure}

Figure \ref{4}(a),(b), and (c) show the CECs for different on-site Coulomb interaction strengths at $\omega$ = -20 meV. We have chosen the range of $U$ to be in between 1.3 and 1.7, because the onset of SDW state in this model occurs near $U \sim 1.2$. Unlike the model due to Ikeda \textit{et al.}, the shape, size and orbital weights are not robust. For instance, there are two concentric  electron pockets around ($0, \pi$) for $U \sim 1.3$, but they disappear as $U$ increases to $\sim$ 1.5 and beyond. Moreover, the central hole pocket at $\Gamma$ is slightly deformed from an elliptical shape. There are nearly semicircular pockets with $d_{yz}$ and $d_{xy}$ character near the central pocket along $k_y = 0$, enabling both intrapocket and interpocket scattering. The pocket around $\Gamma$ deforms into a regular elliptical shape at $U \sim 1.5$ and further transforms into a diamond-like shape at $U \sim 1.7$. On the other hand, the semi-circular pockets along $k_y = 0$ shrink to Dirac nodes near $U \sim 1.5$. These tiny Dirac nodes expand further at $U \sim 1.7$, contributed mainly by $d_{xz}$ and $d_{3z^2 - r^2}$ orbitals, and get attached to the pocket around $\Gamma$.

 Presence of multiple pockets along both the $k_x$ and $k_y$ axes dominated by different orbitals leads to multiple features in the momentum-space QPI pattern for $U \sim 1.3$ corresponding to several scattering vectors (Fig.~\ref{4}(d-f)). As a result of complex overlap of patterns corresponding to different scattering vectors, however, the real space LDOS modulation does not exhibit a clear pattern(Fig.~\ref{4}(g-i)). In near future, with advancement in STM techniques, individual contributions from different orbitals can also be detected.
 Near $U\sim 1.5$, only a few set of scattering vectors are present, as evident from the QPI pattern (Fig.~\ref{4}(e)). Smaller size of elliptical CEC leads to a modulation in real space along $x$ axis $\sim 10a$ slightly larger than what was found in the modal due to Ikeda \textit{et al.}. There is another similarity that the tiny hole pockets do not affect the pattern because of their size and overall orbital weight. The patterns are not very robust, which is again reflected when the intraorbital Coulomb interaction is increased to $U \sim 1.7$ (Fig.~\ref{4}(g-i)). In this case, presence of larger pockets again leads to multiple interpocket and intrapocket scattering vectors as shown in the momentum-space QPI patterns (Fig.~\ref4{f}). As a result, the features of the real-space patterns are relatively weakened while the modulation vector is $\sim 6$a (Fig.~\ref4{i}).

\subsection{Five-orbital model of Kuroki \textit{et al.}}
 
Another five-orbital model reproducing essential features of the electronic structure of IBS is due to Kuroki \textit{et al.}~\cite{kuroki}. This model shares many similarities with the previously discussed two models. However, an important difference to be noted is the presence of hole pocket even around points like $(\pi, \pi)$. For direct comparison, we once again consider the usual unrotated basis of the Fe plane consisting of a single iron atom in the crystallographic unit cell.

\begin{table*}[hbtp]
\scriptsize
	\resizebox{\textwidth}{!}{%
		\begin{tabular}{|p{0.14\linewidth}|p{0.18\linewidth}|p{0.18\linewidth}|p{0.18\linewidth}|}
			\hline
	        {\bf Feature}  &  {\bf Ikeda et al.}  &  {\bf Graser et al.}  &  {\bf Kuroki et al.}  \tabularnewline
			\hline
			\hline
			{\bf hole pocket at $\bm{\Gamma}$}  &  Elliptical hole pocket  with robust shape and size  &  Non-elliptical hole pocket with shape and size highly sensitive to $U$  &  Near elliptical but size moderately sensitive to $U$ \tabularnewline
			\hline
			{\bf Location of Dirac points with respect to Fermi level}  & Robust &  Highly sensitive  & Moderately sensitive  \tabularnewline
			\hline
			\hline
			{\bf Contribution of Dirac pockets in QPI}  & Negligible  &  Negligible  & Negligible  \tabularnewline
			\hline
			\hline
			{\bf Robustness of QPI patterns}  & Robust &  Highly sensitive & Moderately robust  \tabularnewline
			\hline
			{\bf Periodicity of LDOS modulation}  &  Consistent with experimentally observed modulation with robust periodicity $\sim$ $8a$ along AFM direction & Weak modulation along AFM direction, periodicity not robust & Weak modulation along AFM direction, periodicity $\sim 6a - 10a$ \tabularnewline
			\hline
			{\bf Anisotropy in LDOS pattern}  &  Anisotropic but not sufficient enough to explain experimental observations  & Weakly anisotropic, fails to explain experimental observations  &  Weakly anisotropic, fails to explain experimental observations \tabularnewline
			\hline
		\end{tabular}}
        \caption{A comparison of key features of the quasiparticle constant-energy contours together with QPI patterns obtained for the five-orbital models of Graser \textit{et al.}, Ikeda \textit{et al.}, and Kuroki \textit{et al.} as $U$ changes.}  \label{t2}
\end{table*}

\begin{figure}[h]
    \centering
    \includegraphics[scale = 1.0, width = 8cm]{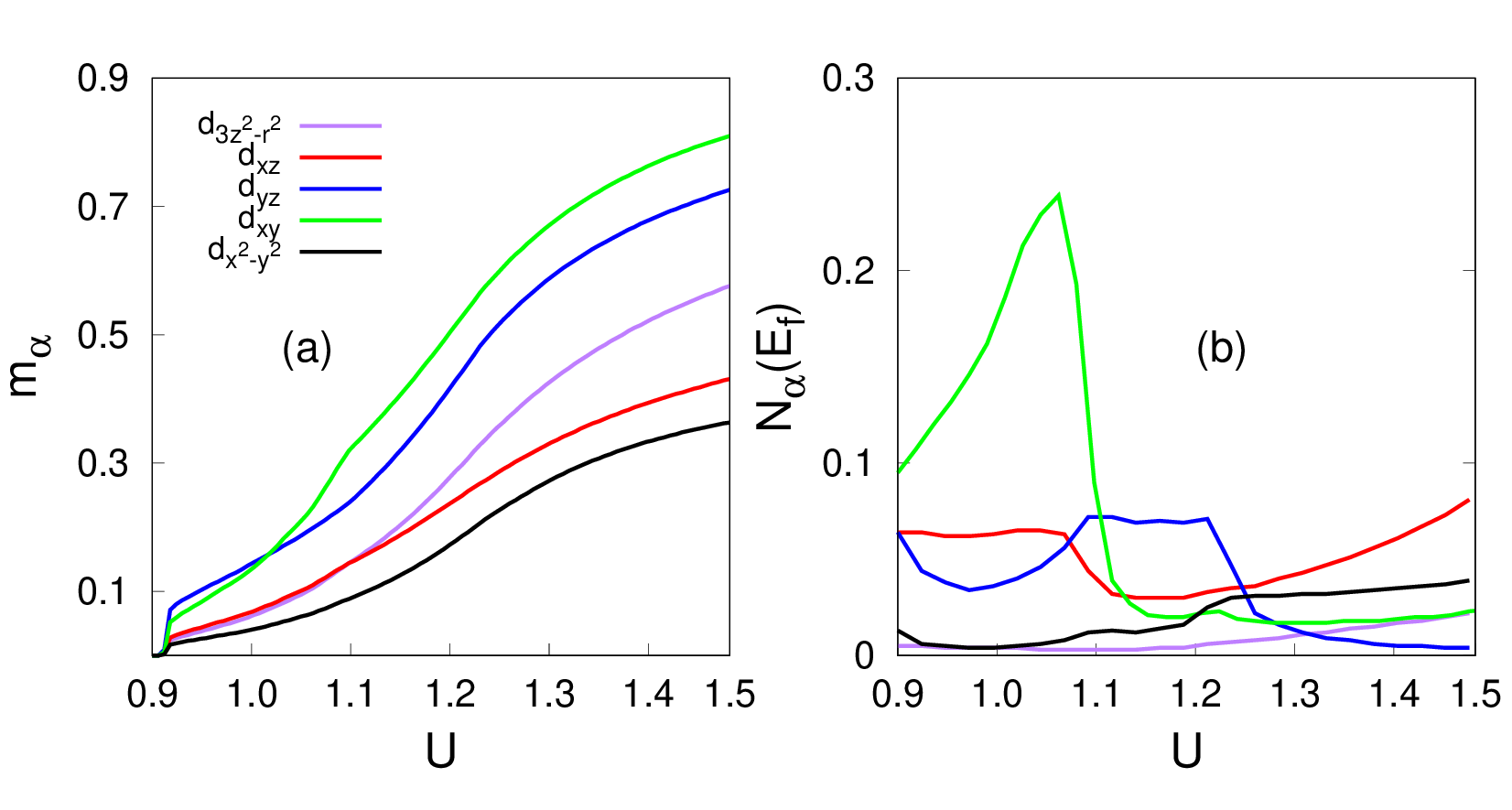}
    \caption{(a) Orbital-resolved magnetization and (b) DOS at the Fermi level as a function of $U$ in the SDW state within the five-orbital model of Kuroki \textit{et al.}}
    \label{5}
\end{figure}

Figure~\ref{5} shows the orbital magnetization and DOS at the Fermi level as a function of $U$. The magnetization trend for individual orbitals mirrors what is found for the model due to Ikeda \textit{et al.} However, the behavior of orbital weights at the Fermi level differs. As shown in Fig.~\ref{5}(b), for smaller values of $U$, the  $d_{xy}$ orbital is the most dominant one and it experiences a sharp rise when $U$ approaches $\sim$ 1.1. The weights of the other dominant orbitals, $d_{xz}$ and $d_{yz}$, remain comparatively smaller, while the rest of the orbitals contributing only insignificantly. The difference between the $d_{xy}$ and $d_{xz}$ weights is more when compared with that in the model due to Ikeda \textit{et al}. As $U$ crosses $\sim$ 1.1, there is a sudden and steep decline in the $d_{xy}$ orbital weight, which is again an indication for the orbital-charge density $n_{xy}$ approaching half filling, and $d_{yz}$ orbital becomes the dominant one, though its weight stays relatively smaller though constant. This pattern persists until approximately $U \sim 1.25$, after which the $d_{xz}$-orbital weight takes the lead in the contribution, with relatively smaller contributions from the other orbitals.

\begin{figure} [h]
    \centering
    \includegraphics[scale = 1.0, width = 8cm]{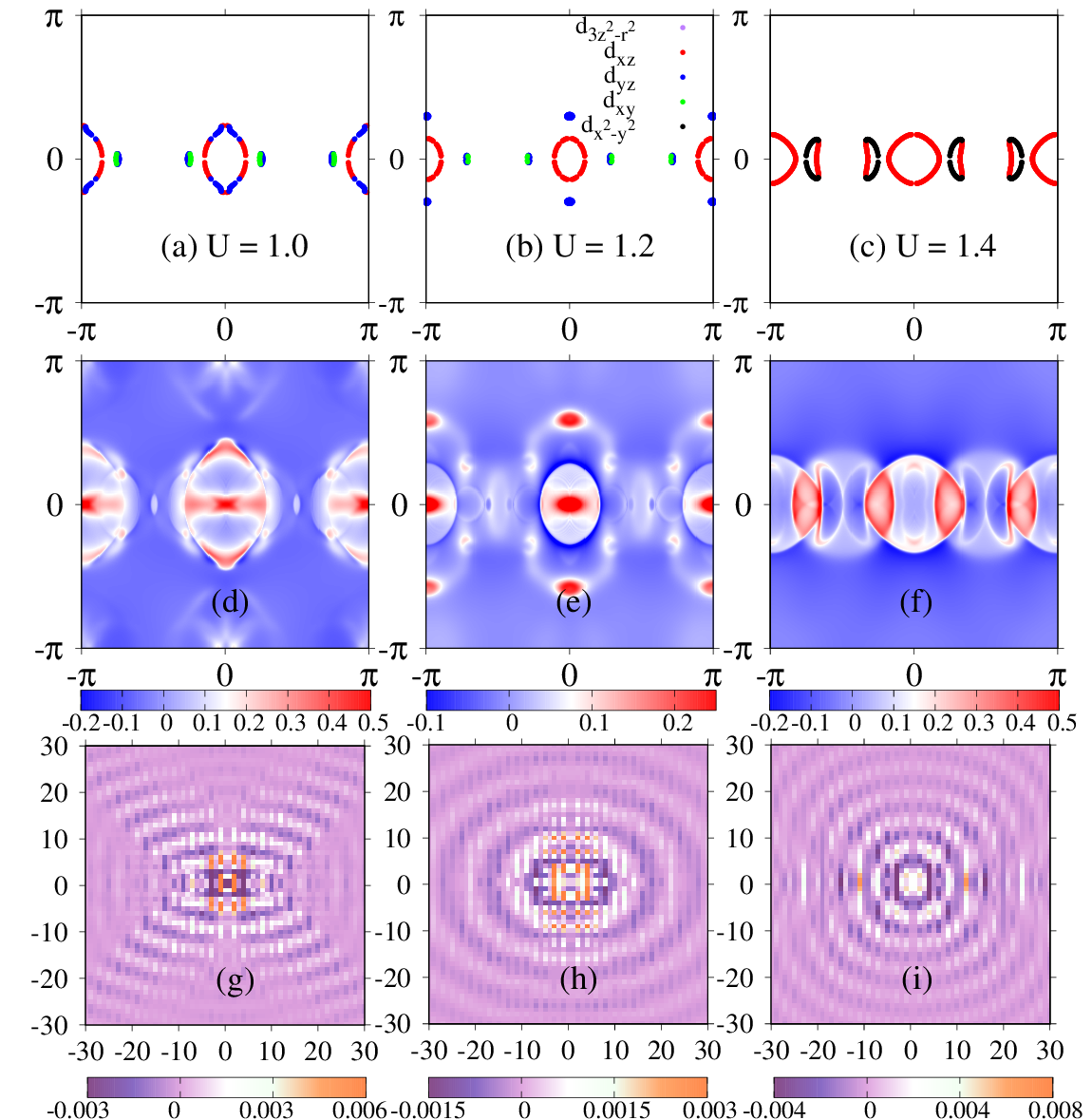}
    \caption{CECs (a-c), QPI patterns (d-f), and real-space LDOS modulations (g-i) for different values of $U$ and $\omega$ = -20 meV within the five-orbital model of Kuroki \textit{et al.}}
    \label{6}
\end{figure}

Figure~\ref{6}(a-c) show CECs for different values of $U$ at $\omega$ = -20 meV. At lower $U$, the CECs has features somewhat similar to those in the model due to Ikeda \textit{et al.}. The CEC consists of a central hole pocket with dominant $d_{xz}$ character, whose shape remains robust with increasing $U$, though it becomes slightly smaller near $U\sim1.2$, afterwards its size increases. The adjacent pockets, initially of $d_{yz}$ and $d_{xy}$ character, shrink as $U$ increases to $1.2$ and transform into crescent-shaped pockets with predominant $d_{xz}$ and $d_{x^2 - y^2}$ character for $U \sim 1.4$. It may also be noted that a pair of Dirac pockets with $d_{yz}$ character lies symmetrically along $k_x = 0$ for $U = 1.2$ and disappears at higher $U$.

A nearly common feature for all $U$ is the largest elliptical pattern around $\Gamma$ as evident in the ${\bf q}$-space  patterns shown in the second row of Fig.~\ref{6}(d-f). For $U \sim 1.0$, the presence of some $d_{yz}$ character in the central hole pocket allows for intraorbital scattering in the diagonal direction between the Dirac electron pocket and the central hole pocket, leading to a strong modulation along a direction with a relative angle of $\pi/4$ with respect to the direction having antiferromagnetic and ferromagnetic arrangements of magnetic moments. Near $U \sim 1.2$, the presence of pockets with the same orbital character along both $k_x$ and $k_y$ results in a weakened overall pattern (Fig.~\ref{6}(h)). However, for $U \sim 1.4$, there is strong scattering  along the antiferromagnetic direction because of the fact that all the pockets are now being dominated by $d_{xz}$ orbital. The scattering between the central hole pocket and the other adjacent pockets results in a modulation of about $ \sim 6a$ units (Fig.~\ref{6}(i)).

\section{Discussion and conclusions}

We have carried out a comparative study of QPI patterns for different  tight-binding models and onsite Coulomb interactions for the SDW state of iron pnictides in order to highlight the importance of interplay between the Coulomb interaction and bandstructure (Table 2). The current work finds that the details of QPI patterns in different models, in addition to being highly sensitive to $U$, may differ in terms of modulation length scale, anisotropy etc. It may further be noted that these patterns are also expected to be sensitive to the Hund's coupling, which was fixed in the current work. The QPI patterns were obtained by considering only those types of impurity atoms which preserved the orbital state of the quasiparticle. Theoretically, it is straightforward to incorporate the interorbital scattering process. The patterns, thus, generated will overlap with those obtained from the intraorbital scattering process, and the resulting patterns will become much more complex, which, however, may be of interest in future with further advancement in STM technique.

We find that robust QPI patterns against a change in $U$ could be obtained only in the model due to Ikeda \textit{et al}. Overall, a small change in the modulation length scale in both momentum as well as real space was observed, which is primarily because of the central role of the larger hole pocket around $\Gamma$, which is dominated by the $d_{xz}$ orbital. A relatively more sensitive behavior is found in the model due to Kuroki \textit{et al.}, and the sensitivity of the QPI patterns to $U$ gets further enhanced in the model due to Graser \textit{et al.}

Earlier, it was proposed that the interpocket scattering between the Dirac pocket lying along the $k_y = 0$ and the larger hole pocket around $\Gamma$ can generate one-dimensional QPI patterns observed in the experiments ~\cite{knolle, dheeraj1, chuang}. However, we find that this does not materialize primarily because of the dominance of different orbitals on the Dirac pocket and the pocket around $\Gamma$. In all the three models, the Dirac pockets are dominated by $d_{xz}$ and $d_{yz}$ orbitals whereas the pocket around $\Gamma$ by $d_{xz}$ orbital. On the other hand, the interpocket scattering between a pair of Dirac pockets along $k_{x}$ could have been expected to result into nearly one-dimensional modulation, however, with $q_x \sim \pi/2$, i.e., $R_x \sim 4a$. Although such a pattern is not found in the model due to Ikeda \textit{et al.} perhaps because of an overlap of patterns generated by intrapocket scattering of the larger pocket around $\Gamma$ and its  interpocket scattering with Dirac pocket. More or less, a similar behavior is noticed in all the other models as well. A minor difference may be noticed for larger $U$ in the model due to Kuroki \textit{et al.}, where the Dirac pocket may also be dominated by $d_{xz}$ orbital. In such a scenario, one could have obtained nearly one-dimensional pattern, however, larger size of the pockets appears to be unfavourable for such a pattern.

To conclude, a nearly one-dimensional modulation in the LDOS depends on the nature of constant-energy contours and the orbital-weight distribution along it, which is highly sensitive to the choice of interaction parameters. Secondly, the presence of Dirac electron pockets near the Fermi level is essential for one-dimensional modulation. However, the periodicity of modulation depends on the pockets involved in the scattering. If the scattering occurs between one of the hole pockets formed around $\Gamma$ and electron pockets formed by Dirac energy bands, then the experimentally observed periodicity of $\sim 8a$ can be seen; otherwise, the periodicity is always less than $8a$, depending on the distance between the pockets involved in scattering.
The extent of the Fermi surface reconstruction and the distribution of orbital weights depend on the strength of on-site Coulombic interaction. Consequently, the quasiparticle interference patterns are also highly sensitive to the choice of interaction parameters, as even minor changes in the interaction parameter strength can alter significantly the details of the spectral function  due to Fermi surface reconstruction in the SDW state, which, in turn, should play an important role in the dimensionality of the modulation and its periodicity.

\section*{ACKNOWLEDGEMENTS} 
D.K.S. was supported through DST/NSM/R\&D\_HPC\_Appl\newline
ications/2021/14 funded by DST-NSM and start-up research grant SRG/2020/002144 funded by DST-SERB.
\\

\end{document}